\documentclass[11pt,a4paper]{llncs}
\usepackage{graphicx}

\usepackage{algorithm}
\usepackage{algpseudocode}

\newcommand{\F}{F}
\newcommand{\Fig}[1]{Fig.~\ref{#1}}

\title{A method for constructing parity-check matrices of non-binary quasi-cyclic LDPC codes} 
 
\author{Stanislav Kruglik, Valeriya Potapova \and Alexey Frolov} 
\institute{Skolkovo Institute of Science and Technology\\Moscow, Russia\\
Institute for Information Transmission Problems\\
    Russian Academy of Sciences\\Moscow, Russia\\
\email{stanislav.kruglik@skolkovotech.ru, valeriya.potapova@skolkovotech.ru, al.frolov@skoltech.ru}}

\begin{document}

\maketitle

\setlength{\arraycolsep}{5pt}

\begin{abstract}
An algorithm for constructing parity-check matrices of non-binary quasi-cyclic low-density parity-check (NB QC-LDPC) codes is proposed. The algorithm finds short cycles in the base matrix and tries to eliminate them by selecting the circulants and the elements of GF(q). Firstly the algorithm tries to eliminate the cycles with the smallest number edges going outside the cycle. The efficiency of the algorithm is demonstrated by means of simulations. In particular, it was shown that NB QC-LDPC codes constructed with use of our algorithm loose less that  0.1 dB in comparison to the best NB LDPC codes.
\end{abstract}


\section{Introduction}
In this paper we consider the problem of constructing parity-check matrices of NB QC-LDPC codes. 

QC-LDPC codes were proposed in \cite{Ta1999},\cite{F}. These codes form an important subclass of LDPC codes \cite{G}, \cite{Ta1981}. These codes also are a subclass of protograph-based LDPC codes \cite{Th}. QC-LDPC codes can be easily stored as their parity-check matrices can be easily described. Besides such codes have efficient encoding and decoding algorithms \cite{KFL}. All of these makes the codes very popular in practical applications. 

NB LDPC codes have advantages over binary LDPC codes. Davey and MacKay \cite{DM} were first who used belief propagation (BP) to decode NB LDPC codes. They showed that NB LDPC codes significantly outperform their binary counterparts. Moreover, non-binary LDPC codes are especially good for the channels with burst errors and high-order modulations \cite{SC}. However we have to mention, that their decoding complexity is still large in comparison to binary LDPC codes \cite{WSM}, \cite{BD}, \cite{DF}.

There are numerous methods for constructing parity-check matrices of binary QC-LDPC codes \cite{ACE}, \cite{PEG}, \cite{IPEG}. In this paper we generalize them to non-binary case.

Our contribution is as follows. An algorithm for constructing parity-check matrices of non-binary quasi-cyclic low-density parity-check (NB QC-LDPC) codes is proposed. The algorithm finds short cycles in the base matrix and tries to eliminate them by selecting the circulants and the elements of GF(q). Firstly the algorithm tries to eliminate the cycles with the smallest number edges going outside the cycle. The efficiency of the algorithm is demonstrated by means of simulations. In particular, it was shown that NB QC-LDPC codes constructed with use of our algorithm loose less that  $0.1$ dB in comparison to the best NB LDPC codes.

\section{Preliminary results}
Consider a binary matrix of size $m \times n$
\[
H_{base} = \left[ h_{i,j}\right] \in \{0,1\}^{m \times n}.
\]
In what follows the matrix will be referred to as the base matrix.

Let us construct a parity-check matrix $H$ of size $ms \times ns$ of NB QC-LDPC code $\mathcal C$. For this purpose we extend the matrix $H_{base}$ with circulant matrices (circulant) multiplied by non-zero elements of GF($q$), i.e. 
\begin{equation}
H =
{ 
\left[
\begin{array}{cccc}
\alpha_{1,1} P_{1,1}  & \alpha_{1,2} P_{1,2}  & \cdots & \alpha_{1,n} P_{1,n}      \\
\alpha_{2,1} P_{2,1}  & \alpha_{2,2} P_{2,2}  & \cdots & \alpha_{2,n} P_{2,n}      \\
\vdots                 & \vdots                 & \ddots & \vdots       \\
\alpha_{m,1} P_{m,1} & \alpha_{m,2} P_{m,2} & \cdots & \alpha_{m,n} P_{m,n}   
\end{array}
\right]},
\end{equation}
where $P_{i,j}$ is a circulant over a binary field of size $s \times s$ and of weight\footnote{the weight of a circulant is a weight of its first row.} $h_{i,j}$, $i = 1, \ldots, m$, $j = 1, \ldots, n$, and $\alpha_{i,j} \in GF(q) \backslash\{0\}$, $i = 1, \ldots, m$, $j = 1, \ldots, n$.

Let us denote the length of the code $\mathcal C$ by $N = ns$, such inequality follows for the rate of the code
$$
R(\mathcal C) \geq1-\frac{m}{n}.
$$

Let $\F$ be some field, by $F[x]$ we denote the ring of all the polynomials with coefficients in $F$. It is well-known that the ring of circulants of size $s \times s$ over $F$ is isomorphic to the factor ring $F^{(s)}[x] = F[x]/(x^s-1)$. Thus with the parity-check matrix $H$ we associate a polynomial parity-check matrix $H(x)$ of size $m \times n$:
\begin{equation}
H(x) = 
\left[
\begin{array}{cccc}
\alpha_{1,1} p_{1,1}(x)  & \alpha_{1,2} p_{1,2}(x)  & \cdots & \alpha_{1,n} p_{1,n}(x)      \\
\alpha_{2,1} p_{2,1}(x)  & \alpha_{2,2} p_{2,2}(x)  & \cdots & \alpha_{2,n} p_{2,n}(x)      \\
\vdots      & \vdots      & \ddots & \vdots          \\
\alpha_{m,1} p_{m,1}(x)  & \alpha_{m,2} p_{m,2}(x)  & \cdots & \alpha_{m,n} p_{m,n}(x)  
\end{array}
\right],
\end{equation}
where $p_{i,j}(x) = \sum\nolimits_{t=1}^{s} {P_{i,j}(t,1)x^{t-1}}$, by $P_{i,j}(t,1)$ we mean an element at the intersection of the $t$-th row and the first column in the matrix $P_{i,j}$. 

\begin{example}
Let us consider the following base matrix
\[
\begin{array}{ccc}
H_{base} = 
\left[
\begin{array}{ccc}
0 & 1 & 1    \\
1 & 0 & 1   \\
\end{array}
\right] 
\end{array}
\]

We can extend it in such a way. Each one in the base matrix is replaced with a polynomial of for $\beta x^z$, where $\beta \in GF(q) \backslash\{0\}$ and $z \in Z_s$, e.g. we have
\[
H(x) = 
\left[
\begin{array}{ccc}
0  & x^2  & 2 x    \\
1 & 0 & 3 x^2   \\
\end{array}
\right].
\]

Note, that $\mathbf{H}(x)$ means the following parity-check matrix
{
\[
\mathbf{H} = 
{
\left[
\begin{array}{c|c|c}
\begin{array}{ccc} 0 & 0 & 0 \\ 0 & 0 & 0  \\ 0 & 0 & 0 \end{array}  & \begin{array}{ccc} 0 & 1 & 0 \\ 0 & 0 & 1  \\ 1 & 0 & 0 \end{array}  & \begin{array}{ccc} 0 & 0 & 2 \\ 2 & 0 & 0  \\ 0 & 2 & 0 \end{array}     \\
\noalign{\vskip 0.1cm} 
\hline 
\noalign{\vskip 0.1cm} 
\begin{array}{ccc} 1 & 0 & 0 \\ 0 & 1 & 0  \\ 0 & 0 & 1 \end{array} & \begin{array}{ccc} 0 & 0 & 0 \\ 0 & 0 & 0  \\ 0 & 0 & 0 \end{array}  & \begin{array}{ccc} 0 & 3 & 0 \\ 0 & 0 & 3  \\ 3 & 0 & 0 \end{array}    \\
\end{array}
\right]}
\]
}
\end{example}

\begin{remark}
We note, that all elements of a circulant are multiplied by the only non-binary value. This is very important for a practical realization of the codes, as the parity-check matrix is very compact and can be stored very efficiently. The proposed construction of the parity check matrix is also good for parallel implementation of the decoding algorithm.
\end{remark}

\section{Description of the algorithm for constructing parity-check matrices of NB QC-LDPC codes}

In this section we provide a description of algorithm for constructing parity-check matrices of NB QC-LDPC codes. The algorithm can be divided into two stages:
\begin{enumerate}
\item Use NB-EXIT analysis to select a base matrix (protograph)
\item Lift the graph using NB-ACE algorithm (proposed below).
\end{enumerate}

NB-EXIT analysis can be done in a similar way as for binary LDPC codes (see detailed description in \cite{Liva}, \cite{Divsalar}). In what follows we propose the lifting method.

\begin{figure}[h]
\centering
\includegraphics[width=0.7\textwidth]{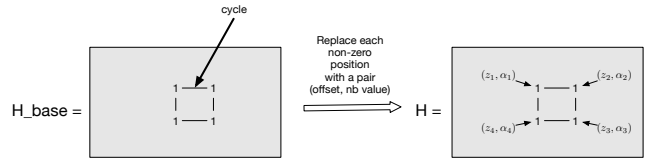}
\caption{Cycle elimination}
\label{1}
\end{figure}

It is well-known, that pseudo codewords are the reason of belief propagation algorithm bad behavior. Pseudo codewords always contain cycles, so the aim of the algorithm is to eliminate short cycles. As an example let us consider the elimination of a cycle of length $4$ (see \Fig{1}).

Let $I$ be subset of rows, containing a cycle, $J$ be subset of columns, containing a cycle. We say that the cycle is eliminated if
\[
\det ({\mathbf{H}(x)_{I,J}}) \ne {\mathbf{0}}.
\]

So the algorithm tries to eliminate as much cycles as possible. If we cannot eliminate a cycle we count the number of edges going from variable nodes in a cycle outside the cycle (so called ACE value of the cycle \cite{ACE}). As there are many cycles and they are of different lengths the algorithm really deals with the ACE vector
\[
ace = (e_4, e_6, \ldots ),
\]
where $e_{2j}$ is a minimal number of edges going outside a cycle of length $2j$ in a parity-check matrix. If there are no cycles of length $2j$, then $e_{2j} = \infty$.

We propose a greedy algorithm, which tries to find an optimal ACE vector on each step. We compare ACE vectors lexicographically, this means, that
\[
(1, 2, 3) < (2, 2, 3), (1, 2, 3) < (1, 3, 3), \ldots
\]

All these stages are described in the algorithm below.

\begin{algorithm}\label{A1}
\caption{Algorithm to construct NB QC-LDPC codes}
\begin{algorithmic}
\State {\bf Input:} $H_{base}$, $Q$, $s$ (size of circulant), depth (maximal cycle length)
\For {$i \in \{1,2, \ldots, n\}$}
	\For {$j \in $ ones in row $i$}
	   \State find all cycles in $H_{base}$ going through variable node $j$ with length $\leq$ depth
	   \For {$test \in \{1,2, \ldots, 100\}$}
	      \State	$H(i,j) \gets (randi(s-1), randi(Q-1))$
          \State	calculate ace vector
	      \If {$ace_{\max}$ $<$ $ace$} $ace_{\max} \gets ace$
          \Else { revert previous value of $H(i,j)$}
	      \EndIf
	   \EndFor
	\EndFor
\EndFor
\State {\bf Output:} $H$ 
\end{algorithmic}
\end{algorithm}

\section{Simulation results}
\begin{figure}[h]
\centering
\includegraphics[width=0.75\textwidth]{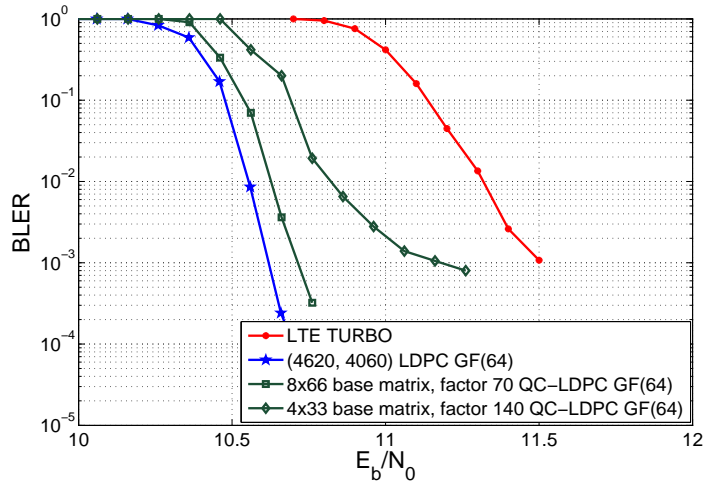}
\caption{Simulation results. Parameters: channel with AWGN, $64$-QAM modulation, $q=64$, code length in field symbols $N=4620$, code dimension in field symbols $K=4060$}
\label{2}
\end{figure}

In this section we present the simulation results. The simulation was carried out in AWGN channel with $64$-QAM modulation. Let $q=64$, $N=4620$ and $K=4060$. At first we add the simulation results for best NB LDPC codes over GF(64), which do not have quasi-cyclic form, and binary Turbo codes. Let us construct two parity-check matrices of NB QC-LDPC codes. In the first case the base matrix has size $4 \times 33$ ($s = 140$), in the second case the base matrix has size $8 \times 66$ ($s = 70$). Obtained results are shown in \Fig{2}.

Note, that a curve for base matrix with size $4\times33$ has a clear error floor. Let us investigate these phenomenon. The reason is the bad minimal code distance. According to \cite{Frolov} the minimal distance of QC-LDPC code has the following upper bound:
\[
D(C) \leq  \lfloor\ell\rfloor!  \ell^{m-\lfloor\ell\rfloor} (m+1),
\]
where $\ell$ is the number of ones in a column (in our case $\ell = 2$) and $m$ is the height of base matrix. 

For the matrix with size $4\times33$ this bound is as follows $D(\mathcal C)\leq 40$. At the same time for the matrix with size $8\times66$ the bound is much better: $D(C)\leq 1152$. The error correcting capabilities of the latter matrix are good, it looses just $0.1$ dB at the level where block error rate (BLER) is equal to $=10^{-3}$.

\section{Conclusion}
Greedy algorithm for constructing parity-check matrices of NB QC-LDPC codes was proposed. The algorithm finds short cycles in the base matrix and tries to eliminate them by selecting the circulants and the elements of GF(q). Firstly the algorithm tries to eliminate the cycles with the smallest number edges going outside the cycle. The efficiency of the algorithm is demonstrated by means of simulations. In particular, it was shown that NB QC-LDPC codes constructed with use of our algorithm loose less that  0.1 dB in comparison to the best NB LDPC codes.

\section*{Acknowledgments}
The research of Stanislav Kruglik was supported by the
Russian Foundation for Basic Research (project 16-01-00716).


\begin{thebibliography}{9}

\bibitem{Ta1999}
R.~M.~Tanner.
\newblock  On quasi-cyclic repeat-accumulate codes.
\newblock in \emph{Proc. 37th Allerton Conf. Commun., Contr., Comput.}, Monticello, IL, Sep.
22--24, 1999, pp. 249--259, Allerton House.

\bibitem{F}
M.~P.~C.~Fossorier.
\newblock Quasi-cyclic low-density parity-check codes from circulant permutation matrices.
\newblock \emph{IEEE Trans. Inf. Theory},
vol. 50, no. 8, pp. 1788--1793, Aug. 2004.

\bibitem{G}
R.~G.~Gallager.
\newblock \emph{Low-Density Parity-Check Codes}.
\newblock Cambridge, MA: M.I.T. Press, 1963.

\bibitem{Ta1981}
R. M. Tanner.
\newblock A recursive approach to low-complexity codes.
\newblock \emph{IEEE Trans. Inf. Theory}, vol. 27, no. 5, pp. 533--547, Sep. 1981.

\bibitem{Th}
J.~Thorpe.
\newblock Low-density parity-check (LDPC) codes constructed from protographs.
\newblock JPL, IPN Progress Rep., Aug. 2003, vol. 42--154.

\bibitem{CZLF}
Z.~Li, L.~Chen, L.~Zeng, S.~Lin, and W.~H.~Fong.
\newblock Efficient encoding of quasi-cyclic low-density parity-check codes.
\newblock \emph{IEEE Trans. Commun.},
vol. 54, no. 1, pp. 71--78, Jan. 2006.

\bibitem{KFL}
F.~R.~Kschischang, B.~J.~Frey, and H.-A.~Loeliger.
\newblock Factor graphs and the sum-product algorithm.
\newblock \emph{IEEE Trans. Inf. Theory}, vol. 47, no. 2, pp. 498--519, Feb. 2001.

\bibitem{DM}
M. Davey and D. MacKay, Low-density parity check codes over GF(q), IEEE Commun. Lett., vol. 2, no. 6, pp. 165--167, Jun. 1998

\bibitem{SC}
H. Song,  J. R. Cruz, Reduced-Complexity Decoding of Q-ary LDPC Codes for Magnetic Recording, IEEE Trans. on Magnetics, vol. 39, no. 2, March 2003

\bibitem{WSM}
H. Wymeersch, H.Steendam, M.Moeneclaey, Log-domain decoding of LDPC codes over GF(q), IEEE Int. Conf. on Communications, 2004, pp 772--776

\bibitem{BD}
L. Barnault and D. Declercq, Fast Decoding Algorithm for LDPC over GF($2^q$), The Proc. 2003 Inform. Theory Workshop, Paris, France, pp. 70--73, Mar. 2003

\bibitem{DF}
D.Declercq, M. Fossorier, Decoding Algorithms for Nonbinary LDPC Codes over GF(q), IEEE Trans. On Communications, vol.55, no.4, April 2007, pp 633--643


\bibitem{ACE} \emph{D. Vukobratovic and V. Senk}, Generalized ACE Constrained Progressive Edge-Growth LDPC Code Design, in IEEE Communications Letters, vol. 12, no. 1, pp. 32-34, January 2008.

\bibitem{PEG}
H.Xiao, A. H. Banihashemi, Improved Progressive-Edge-Growth (PEG) Construction of Irregular LDPC Codes, IEEE Comm. Letters, vol.8, no. 12, December 2004, pp.715--717

\bibitem{IPEG}
X. Hu, E. Eleftheriou, D. Arnold, Irregular Progressive Edge-Growth (PEG) Tanner Graphs, ISIT 2002, Lausanne, Switzerland, June 30--July 5, 2002

\bibitem{Liva}
G. Liva and M. Chiani.
\newblock Protograph LDPC Codes Design Based on EXIT Analysis.
\newblock IEEE GLOBECOM 2007 - IEEE Global Telecommunications Conference, Washington, DC, 2007, pp. 3250--3254.

\bibitem{Divsalar}
T. Y. Chen, K. Vakilinia, D. Divsalar and R. D. Wesel. 
\newblock Protograph-Based Raptor-Like LDPC Codes.
\newblock \emph{IEEE Trans. on Comm.}, 
vol. 63, no. 5, pp. 1522--1532, May 2015.

\bibitem{Frolov} \emph{A. Frolov}, Upper Bounds on the Minimum Distance of Quasi-Cyclic LDPC codes // XIV International Workshop on Algebraic and Combinatorial Coding Theory (ACCT 2014) September 7-13, 2014, Kaliningrad, Russia. P. 163--168.

\end{thebibliography}
\end{document}